\documentclass[angeo]{copernicus}
\usepackage{amsthm}
\pdfoutput=1

\begin{document}

\title{Polyphase alternating codes}

\author[1]{Markku Markkanen}
\author[2]{Juha Vierinen}

\affil[1]{Eigenor Corporation}
\affil[2]{Sodankyl\"a Geophysical Observatory}

\runningtitle{Polyphase alternating codes}
\runningauthor{Markkanen and Vierinen.}
\correspondence{Juha Vierinen\\ (j@sgo.fi) }

\received{}
\pubdiscuss{} 
\revised{}
\accepted{}
\published{}

\firstpage{1}

\maketitle

\begin{abstract}
This work introduces a method for constructing polyphase alternating
codes in which the length of a code transmission cycle can be $p^m$ or
$p-1$, where $p$ is a prime number and $m$ is a positive integer. The
relevant properties leading to the construction alternating codes and
the algorithm for generating alternating codes is described. Examples
of all practical and some not that practical polyphase code lengths
are given.
\end{abstract}


\introduction

Alternating codes \citep{lehtinenPhd} are widely used in incoherent
scatter radar measurements and their properties are well
known. Currently, the only so called type 1 code lengths are powers of
two, which sometimes causes inflexibility when designing radar
experiments. With this in mind, \cite{sulzer93} proposed a new type of
alternating codes (type 2) that made it possible to use other code
lengths. These codes also produce unambiguous back-scatter
autocorrelation function estimates of the target. But there does not
exist a search strategy efficient enough for finding longer type 2
codes. To best of our knowledge, the longest type 2 alternating code
length is 14, which is not necessarily large enough for all
applications.

\begin{figure}[t]
\begin{center}
\includegraphics[width=\columnwidth]{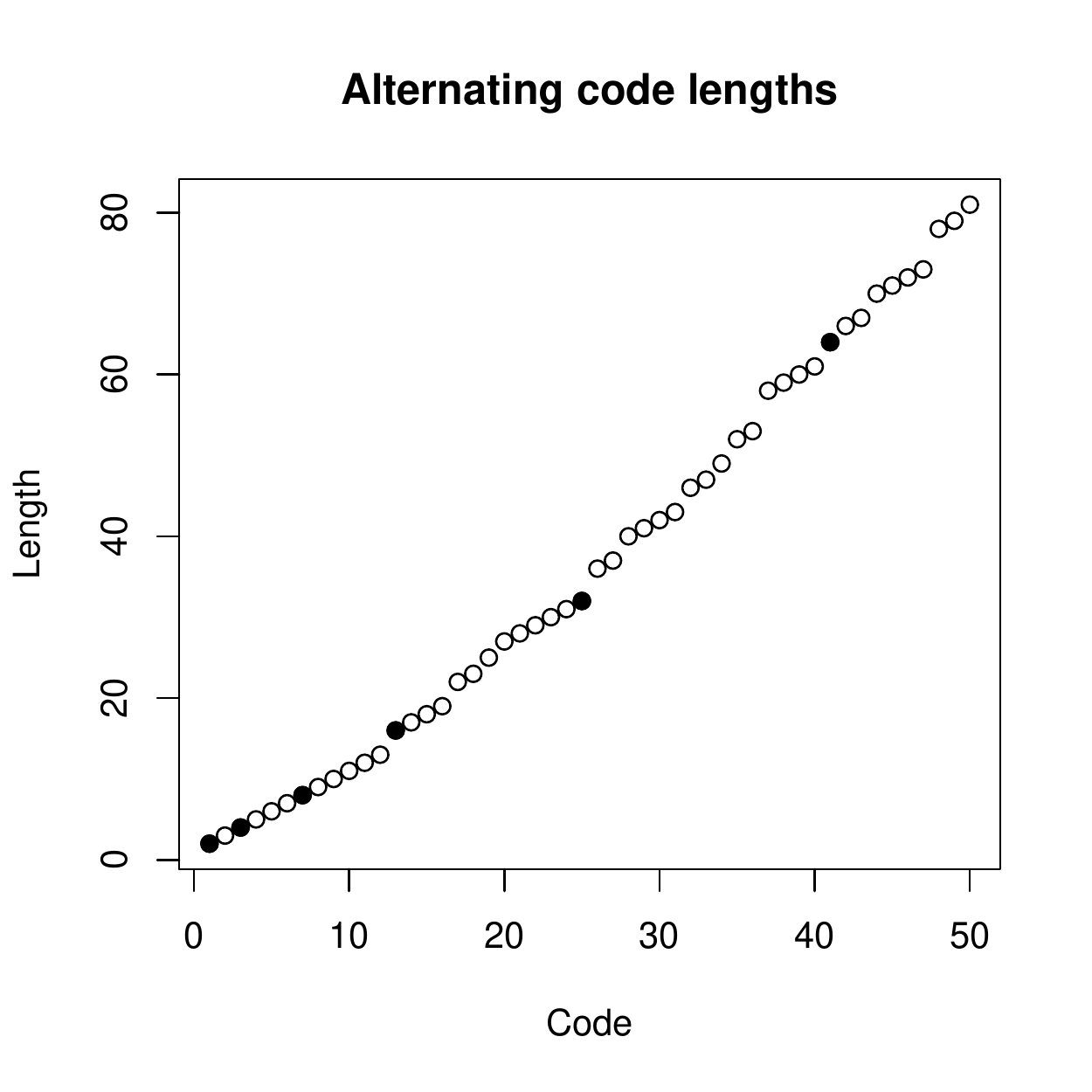}
\end{center}
\caption{First 50 alternating code lengths, solid black circles
represent known binary phase alternating codes, open circles represent new
polyphase alternating code lengths.}
\label{lengths}
\end{figure}

In this paper we generalize previous work \citep{markkanen97} and
apply it to polyphase codes, i.e., codes that are phase coded with two
or more different phases. We have identified two different types of
polyphase alternating codes. The first type includes codes that have
$p$ phases and code lengths of $p^m$, where $p$ is a prime number and
$m$ is a positive integer. The other class contains codes with $p-1$
phases and code length $p-1$. The number of codes in these kinds of
alternating code sets is the same as the number of bauds in a
code. The first $50$ alternating code lengths are shown in Fig.
\ref{lengths}.

This paper concentrates on polyphase alternating codes satisfying the
so called weak condition. Strong codes can be generated from weak
codes using the method described by \cite{sulzer89}. This method also
works for the polyphase alternating codes presented in this work. To
create an alternating code set satisfying the strong condition, the
code set is duplicated. For each code in the duplicated part, bauds
with even indices are multiplied by $-1$.


\section{Construction of $p$-nary alternating codes}

By a $p$-nary radar pulse we mean a phase modulated pulse, where the
complex phases of bauds belong to set $E_p
=\{1,\alpha,\dots,\alpha^{p-1}\}$, where $\alpha = e^{2\pi i/p}$. In
what follows, $p$ is a prime number unless otherwise stated. A
$p$-nary alternating code of length $n$ is a set of sequences $A_0 =
(a_{0,0},\dots,a_{0,n-1}),\dots,A_{n} = (a_{n-1,0},\dots,a_{n-1,n-1})$
of $p$-nary pulses satisfying the natural extension of the \emph{weak
condition} given in \cite{lehtinenPhd} for binary codes:

\newtheorem{condition}{Condition}

\begin{condition}
\label{condLeht} 
For each $i,i',j$ and $j'$ where $j-i = j'-i',\ i\ne j,$ and $i\ne i'.$ 

\[
\sum_{k=0}^{n} a_{k,i} \overline{a_{k,j}} \overline{a_{k,i'}  \overline{a_{k,j'}}} = 0 
\]

\end{condition}

Let us denote by $M$ the operator, which shifts the elements of a
sequence cyclically by one, that is

\[
M\,(a_0,\dots,a_{n-1}) = (a_1,\dots,a_{n-1},a_0). 
\]

Let $A = (a_0,\dots,a_{n-1})$ be a $p$-nary sequence and let us denote
by $U$ the unit sequence $U = (1,1,\dots,1)$. By multiplication of two
sequences we mean pointwise multiplication of them and will denote it
by $\otimes$. Similarly we will mean by conjugate of $A$ the pointwise
complex conjugate of $A$ and denote it by $\overline A$. Then $A
\otimes \overline A = U$, because $|a_i| = 1$ for all $i$. 

Let us denote by $C$ the set of sequences $A, MA,\dots,M^{n-1}A,
U$. Here $M^i$ means $i$ repeated cyclical shift operations. We will
show that when $A$ satisfies the following two conditions, $C$ will
constitute a $p$-nary alternating code set:

\begin{condition}
\label{alldifferent}
All the sequences of $C$ are different.
\end{condition}

\begin{condition}
\label{permutes}
Multiplying all the sequences of $C$ by a fixed one
of it permutes $C$.
\end{condition}

To prove this, we need the following properties of the sequences
satisfying conditions \ref{alldifferent} and \ref{permutes}:

\newtheorem{property}{Property}

\begin{property}
\label{prop1}
The sum $s$ of elements of $A$ is $-1$.
\end{property}

This can be seen by considering the sequence $S =
A+MA+\cdots+M^{n-1}A+U$, which equals $(s+1)U$ (the sequences
$A,\dots,M^{n-1}A$ contain in each location all the elements of $A$ in
some cyclic order). Because by condition \ref{permutes} multiplication by $A$
permutes the terms of $S$, $A\otimes S = S$, and because $A\otimes U =
A$, it follows that

\[
(s+1)A = (s+1)U.
\]

Now condition \ref{alldifferent} requires that $A\ne U$. Thus $s+1 =
0$ and $s = -1$.

Notice that in the case of $p = 2$, when $E_p = \{ 1,-1\}$, it follows
from property \ref{prop1} that n is odd.

\begin{property}
\label{conj}
There exists $i_0$ so that $\overline {M^i A} = M^{i+i_0} A$
for all $i$.
\end{property}

By condition \ref{permutes} there is $i_0$ so that $ A \otimes M^{i_0}
A = U$, that is, $M^{i_0} A = \overline A$. Then

\[
\overline {M^i A} = M^i \overline A = M^i M^{i_0} A = M^{i+i_0}
A,\quad {\rm for\ any\ }i. 
\]

Let as denote by $A^i$ the sequence formed by taking the the $i^{th}$
elements of sequences $A_0,\dots, A_{n-1}$.  Condition \ref{condLeht}
is then the requirement that sum of elements of sequence $D = A^i
\otimes \overline{A^j} \otimes \overline {A^{i'} \otimes
\overline{A^{j'}}}$ added by $1$ (coming from $A_n = U$) is 0, that is
sum of elements of $D$ is $-1$.

Now $A^i = M^i A$ and thus

\[
D = M^iA \otimes \overline{M^jA} \otimes \overline {M^{i'}A \otimes
  \overline{M^{j'}A}} .
\]

Property \ref{conj} says then that $D \in C$ and if we can show that
$D \ne U$, it follows from property \ref{prop1} that the sum of elements
of $D$ is indeed $-1$.

Because $B_1 \otimes \overline {B_2} \ne U$ for any $B_1 \ne B_2$ it
is enough to show that

\begin{equation}
\label{eqdiff}
M^iA \otimes \overline{M^jA} \ne M^{i'}A \otimes \overline{M^{j'}A} .
\end{equation}

Now the left hand side is $M^iA \otimes M^{j+i_0} A = M^i (A \otimes
M^{j-i+i_0} A)$ and similarly the right hand side is $M^{i'} (A
\otimes M^{j'-i'+i_0} A)$. Because $j-i = j'-i'$, the sequences inside
parentheses are equal, and because $i\ne j$ they are not equal to
$U$. Then indeed $M^i (\dots) \ne M^{i'}(\dots)$, because $i\ne i'$.

We will finally make the number of codes and the length of them equal
by copying the first element of each sequence the end of the
sequence. Then $A^n = A = M^n A$.

To see that condition \ref{condLeht} is still satisfied, let us first suppose that $n$ is in the set $\{i,j,i',j'\}$ but $0$ is not. Then that condition follows trivially from what was said above (by dropping the first
element of each sequence we have the same codes, only in order $M
A,\dots, M^{n-1}A, A, U$). If both $0$ and $n$ belong to the set
$\{i,j,i',j'\}$, the only possibility is $i = 0$, $j = i' = n/2$. Then
$n$ must be even, so we can suppose that $p \ne 2$. With the above
choice of indices the left hand side of Eq. \ref{eqdiff} is $A
\otimes \overline {M^{n/2} A}$ and the right hand side is $M^{n/2} A
\otimes \overline A$. As they are conjugates, they are unequal, unless
they are both equal to $U$. But $M^{n/2} A \ne A$ and thus $A \otimes
\overline {M^{n/2} A} \ne A \otimes \overline {A} = U$, and so $D \ne
U$, meaning that condition \ref{condLeht} is satisfied also in this
case.

The requirement that $p$ is prime is necessary for this method of
constructing alternating codes, with composite $p$ there isn't any
sequences $A$ satisfying conditions \ref{alldifferent} and
\ref{permutes} (unless the elements of $A$ belong already to $E_p'$ for some prime number $p'$ dividing $p$). The essential reason for that is that for composite $p$ sequences $A, A\otimes A, A\otimes A\otimes A,\dots$ can have different number of ones.

\section{Construction of sequence $A$}
\label{constructSect}

Instead of set $E_p$ we will consider the set $\mathbb{Z}_p = \{
0,\dots,p-1\} \bmod p$ of possible exponents of $\alpha$. Instead of
sequences with elements in $E_p$ we then have vectors in vector space
$\mathbb{Z}_p^n$, multiplication of sequences corresponds to addition,
multiplication by complex conjugate corresponds to subtraction and $U$
is replaced by 0-vector ${\bf 0} \in \mathbb{Z}_p^n$. The operator $M$
corresponds to a linear operator mapping each base vector of
$\mathbb{Z}_p^n$ to the previous one, except the first one, which is
mapped to the last one.

Let us thus have a vector $A = (a_0,\dots,a_{n-1})\in \mathbb{Z}_p^n$
and set $C = \{A, MA,\dots,M^{n-1}A,{\bf 0}\}\subset
\mathbb{Z}_p^n$. Conditions \ref{alldifferent} and \ref{permutes} then
correspond to the following conditions:

\newtheorem*{conda}{Condition 2a}
\newtheorem*{condb}{Condition 3a}

\begin{conda}
All the vectors in $C$ are different.
\end{conda}

\begin{condb}
Adding a fixed vector of $C$ to all vectors belonging to it permutes
$C$.
\end{condb}

The sum of any two vectors in $C$ belong to $C$ by condition
\ref{permutes}a. In fact, when condition \ref{alldifferent}a is
satisfied, condition \ref{permutes}a is equivalent to the condition that sum of any two vectors in $C$ belong to $C$, because addition of a fixed vector is then bijection. Because for any $B\in C$ also $2B = B + B \in C$ and similarly for any $k \in \mathbb{Z}_p$ by induction, it follows that $C$ is a vector subspace of $\mathbb{Z}_p^n$. If $\dim(C) = m$, number of vectors in $C$ is $p^m$, and thus $n = p^m -
1$.

Let us now consider in $\mathbb{Z}_p$ a $m^{th}$-order difference equation

\begin{equation}
\label{diffeq}
x_{i+m} =  b_{m-1} x_{i+m-1} + \dots + b_0x_i,\quad i = 0,\dots,
\end{equation}
with $b_0,\dots,b_{m-1} \in \mathbb{Z}_p$.
Because there is only a finite number of different m-tuples of elements
of $\mathbb{Z}_p$, any sequence, which is solution of (\ref{diffeq}), is periodic
(possible beginning with a non-periodic part if $b_0 = 0$). Let us
suppose that (\ref{diffeq}) has a solution $A' = \{
a_i\}_{i=0}^\infty$ with period $n = p^m-1$. Then the vector $A =
(a_0,\dots,a_{n-1})\in \mathbb{Z}_p^n$ satisfies conditions \ref{alldifferent}a
and \ref{permutes}a. 

Periodic part of $A'$ contains all $p^m-1$ different non-zero
m-tuples, and so $A'$ can't have non-periodic start. Then $a_n =
a_0,a_{n+1} = a_1,\dots$, and the m-tuples $(a_0,\dots,a_{m-1})$,
$(a_1,\dots,a_{m})$, $\dots,(a_{n-1},\dots, a_{n+m-2}),(0,\dots,0)$,
which are the starts of vectors $A,MA,\dots,$ $M^{n-1}A, {\bf 0}$ are
all the $p^m$ different m-tuples of $\mathbb{Z}_p$, proving condition 1a. If
$B_1,B_2 \in C$, there is $B_3 \in C$ such that the first $m$ elements
of $B_1+B_2$ are equal to the first $m$ elements of $B_3$. Because
$B_1+B_2$ is also a solution of (\ref{diffeq}), it follows that
$B_1+B_2 = B_3$, proving condition 2a. 

Notice that we can choose as our $A$ any nonzero vector of $C$,
e.g., one starting with $m-1$ zeros followed by $1$.

Thus, in order to satisfy the original condition for alternating
codes, it is sufficient that the following condition is satisfied:

\begin{condition}
The $m^{th}$ order difference equation in $\mathbb{Z}_p$ has a period $p^m - 1$
\end{condition}

\section{The number of $p^m$-type alternating codes}

If all the roots of polynomial $Q(x) = x^m-b_{m-1}x^{m-1}-\dots - b_0$
are different, the general solution of the difference equation
(\ref{diffeq}) is

\[
x_i = c_1\alpha_1^i+\dots+c_m\alpha_m^i,\quad i = 0,\dots, 
\]
where $\alpha_1,\dots,\alpha_m$ are the roots of $Q(x)$ and
$c_1,\dots,c_m$ are arbitrary coefficients. If $Q(x)$ is irreducible
in $\mathbb{Z}_p[x]$ and there is no integer $d$ smaller than $n$ such
that $Q(x)$ divides $x^d-1$, the roots of $Q(x)$ are different and the
period of all sequences $\{ 1,\alpha_i,\alpha_i^2,\dots\}$ is $n$, and
so is then also the period of sequence $\{ x_0,x_1,\dots\}$. Thus each
$m$-degree polynomial $Q(x)\in \mathbb{Z}_p[x]$ satisfying the above
mentioned conditions determines a $p$-nary alternating code, and it is
easy to see that different $Q$:s determine different codes by
comparing for all locations of 1, the elements following $m$-tuple
$(0,\dots,0,1,0,\dots,0)$ in the solutions of corresponding difference
equations.

It is possible to show \citep[e.g.,][p.~85]{lidl97} that the number of
different polynomials $N_c$ satisfying the above mentioned conditions
is

\[
N_c = \frac{\varphi(p^m-1)}{m}. 
\]

Here $\varphi(i)$ is the Euler $\varphi$-function, which is the number
of integers smaller than $i$ which do not have a common factor with
$i$. It can also be shown that there are no other sets $C$ satisfying
conditions 1a and 2a, so the number $N_c$ given above is the number of
different alternating codes satisfying those conditions. This means especially that for any prime number $p$ and any positive integer $m$ there exist $p$-nary alternating codes of length $p^m$.

\section{$p$-nary alternating codes of length $p$}
\label{thebeef}

We will now look more closely at the case $m = 1$. In this case the difference equation is simply

\begin{equation}
\label{diffeq1}
x_{i+1} = b_0 x_i\quad b_0 \in \mathbb{Z}_p,\quad i = 0,1,\dots \quad ,
\end{equation}
and if we choose $x_0 = 1$, it's solution is $x_i = b_0^i$ $(\bmod\,
p)$. We get thus a suitable $A$, if and only if all the numbers
$1,b_0,\dots,b_0^{p-2} \bmod p$ are different, that is, if $b_0$ is
the generator of the cyclic multiplicative group $\mathbb{Z}_p^{\ast}$.

As an example, let us have a look at primes 3 and 5.
The only generator of $\mathbb{Z}_3^{\ast}$ is $b_0 = 2$, giving the set of 
exponent sequences 

\[
C = \{(1,2,1),
(2,1,2),
(0,0,0)\}, 
\]
for the base element $\alpha = e^{2\pi i/3}$. For $\mathbb{Z}_5^{\ast}$ there are two generators $b_0 = 2, 3$, giving the sets

\begin{eqnarray}
C & = \{&(1,2,4,3,1), (2,4,3,1,2), (4,3,1,2,4),\nonumber \\
  &   &  (3,1,2,4,3), (0,0,0,0,0)\,\}\nonumber
 \end{eqnarray}
and

\begin{eqnarray}
C & = \{&(1,3,4,2,1), (3,4,2,1,3), (4,2,1,3,4),\nonumber  \\
  &   & (2,1,3,4,2),(0,0,0,0,0)\,\}\nonumber
\end{eqnarray}
for the base element $\alpha = e^{2\pi i/5}$. 

Let us suppose that $b_0$ is a generator of $\mathbb{Z}_p^{\ast}$ and
$A$ is the solution of the corresponding difference equation
(\ref{diffeq1}). It was shown in section \ref{constructSect} that the
sequence $kA$ belongs to the code for all $k=0,\dots,p-1$, and by
changing the order of sequences we thus have $C = \{{\bf
0},A,2A,\dots,(p-1)A\}$. This means that the columns of the
alternating code are suitably chosen columns of the unnormalized
Fourier matrix $F_p = (f_{ij})$ with $f_{ij} = e^{ij2\pi
\sqrt{-1}/p}$.

The columns of $F_p$ are orthogonal and form a closed set under
pointwise multiplication. This is true for arbitrary $p$ and we can
use them, or rather the columns of corresponding exponents for
searching $p$-nary alternating codes for any $p$, analogously to the
use of Walsh sequences in \cite{lehtinenPhd}.

Because all the pointwise products of columns of $F_p$ and their
conjugates are also columns of $F_p$, set $C$ is an alternating code
if the pointwise products of it's columns corresponding indices of
condition \ref{condLeht} are different from $U$. As was said earlier,
multiplication by complex conjugate corresponds subtraction of
exponents, and then for arbitrary $p$ the set $C = \{{\bf
0},A,2A,\dots,(p-1)A\}$ with the defining sequence $A =
(a_0,\dots,a_{n-1})\in \mathbb{Z}_p^n$ is an alternating code if $A$
satisfies the following condition (analogous to the condition for Walsh indices given in \cite{lehtinenPhd})

\begin{condition}
\label{condLehtA} 
For each $i,i',j$ and $j'$ where $j-i = j'-i',\ i\ne j,$ and $i\ne i',$ 

\[
(a_i - a_j) - (a_{i'} - a_{j'}) \ne 0,
\]

\end{condition}

The above condition can be rephrased as:
all differences ($\bmod\ p$) of values $a_i,a_j$ of elements of $A$ with fixed difference of indices $i,j$ are different.

We will now consider again for a prime $p$ the sequence $A =
\{1,b_0,\dots,b_0^{p-2}\} \bmod p$, which defines an alternating code
of length $p-1$ (we drop away the duplicate element from the end).

Because of periodicity of $b_0^i$ the sequence A satisfies an even
stronger condition: all differences ($\bmod\ p$) of values $a_i,a_j$
of elements of $A_1$ with fixed difference ($\bmod\ p-1$) of indices
$i,j$ are different.

Then it trivially satisfies also the following condition: all
differences ($\bmod\ p-1$) of indices $i,j$ of elements of $A$ with
fixed difference ($\bmod\ p$) of values $a_i,a_j$ are different.

For the ``dual'' sequence $A'$ with indices $1,b_0,\dots,b_0^{p-2}$
and corresponding values $0,1\dots,p-2$ this means that all
differences ($\bmod\ p-1$) of values of elements of $A'$ with fixed
difference ($\bmod\ p$) of indices are different.

But this is again stronger than the condition for $(p-1)$-nary alternating codes:
all differences ($\bmod\ p-1$) of values of elements of $A'$ with fixed difference of indices are different,
and thus the sequence $A'$ defines a $(p-1)$-nary alternating code of length $p-1$!

{\bf Example:} dual 7-nary and 6-nary sequences:

When $p = 7$ and $b_0 = 3$ , we have the following situation:

{\begin{tabular}{cccccccc}
  $\mathbb{Z}_7$ &&&&&&& $\mathbb{Z}_6$\\
  index: & 0 & 1 & 2 & 3 & 4 & 5 & :value\\
  value: & 1 & 3 & 2 & 6 & 4 & 5 & :index\\
\end{tabular}

\noindent This gives sequence $\{1,3,2,6,4,5,1\}$ defining a 7-nary alternating code and sequence $\{0,2,1,4,5,3\}$ defining a 6-nary alternating code.

It can be noted that whereas in $p$-nary codes the first and last
columns are identical and the constant column of $F_p$ is not in the
code, in $(p-1)$-nary codes all the columns of $F_{p-1}$ are in the code
once.

The number of $p$-nary alternating codes of length $p$ is $\varphi(p-1)$
and one could think that it is also the number of $(p-1)$-nary
alternating codes of length $p-1$. However, this is not the case,
instead different $p$-nary codes give only different sequences of the
same $(p-1)$-nary code, and thus the construction presented here gives
only one $(p-1)$-nary alternating code for each prime $p$. This can be seen by noticing that if $b_0$ and $b_1$ are generators of $\mathbb{Z}_p^{\ast}$, there is $k_0$ such that $b_0 = b_1^{k_0}$. If then $A = \{a_1,a_2,\dots\}$ and $A' = \{a'_1,a'_2,\dots\}$ are dual sequences of $\{1,b_0,\dots\}$ and $\{1,b_1,\dots\}$,

\[
a'_{b_0^k} = a'_{b_1^{k_0k}} = k_0 k = k_0 a_{b_0^k}, 
\]
meaning that $A' = k_0A$.

\section{Algorithm for generating $p^m$-length alternating codes}

Finding alternating codes is a fairly simple computation that involves
going through all possible m-tuples $b = (b_0,...,b_{m-1})$ with
elements in $\mathbb{Z}_p$. This involves going through only $p^m$
alternatives. For each $b$, we check if the corresponding $m^{th}$
order difference equation (\ref{diffeq}) has a period of $p^m - 1$
(notice that the equation is calculated in modulo $p$ arithmetic). If
this is true, then the difference equation with coefficients $b$ is a
generator for an alternating code set, which can then be defined as

\begin{eqnarray}
& A_j & =  \{ \alpha^{x_{j+m+1}},...,\alpha^{x_{j + p^m + m}} \}, \quad j \in
{1,...,p^m-1},\nonumber  \\
& A_{p^m} & =  \{ 1,1,...,1 \}, \nonumber
\end{eqnarray}
where $\alpha = e^{2\pi i/p}$ and $x_i$ are the values of the
generating difference equation (\ref{diffeq}), using initial values
$x_1,...,x_{m-1} = 0$ and $x_{m}=1$.

To generate a $(p-1)$-nary code, one first generates the corresponding
$p$-nary code and then forms the code group using the transformation
described in Sect. \ref{thebeef}.

Table \ref{coderes} lists alternating codes up to code length 366,
only one code per code length is listed.  The codes are expressed in
terms of number of phases $p$, generating coefficients $b =
(b_0,...,b_{m-1})$ and code length. As an example, a 25 baud
alternating code set is shown as phases in Fig. \ref{code25fig}. A
program for generating weak and strong polyphase alternating codes is
available at request from the authors.

\begin{figure}[t]
\begin{center}
\includegraphics[width=\columnwidth,viewport=1 0 413 431,clip]{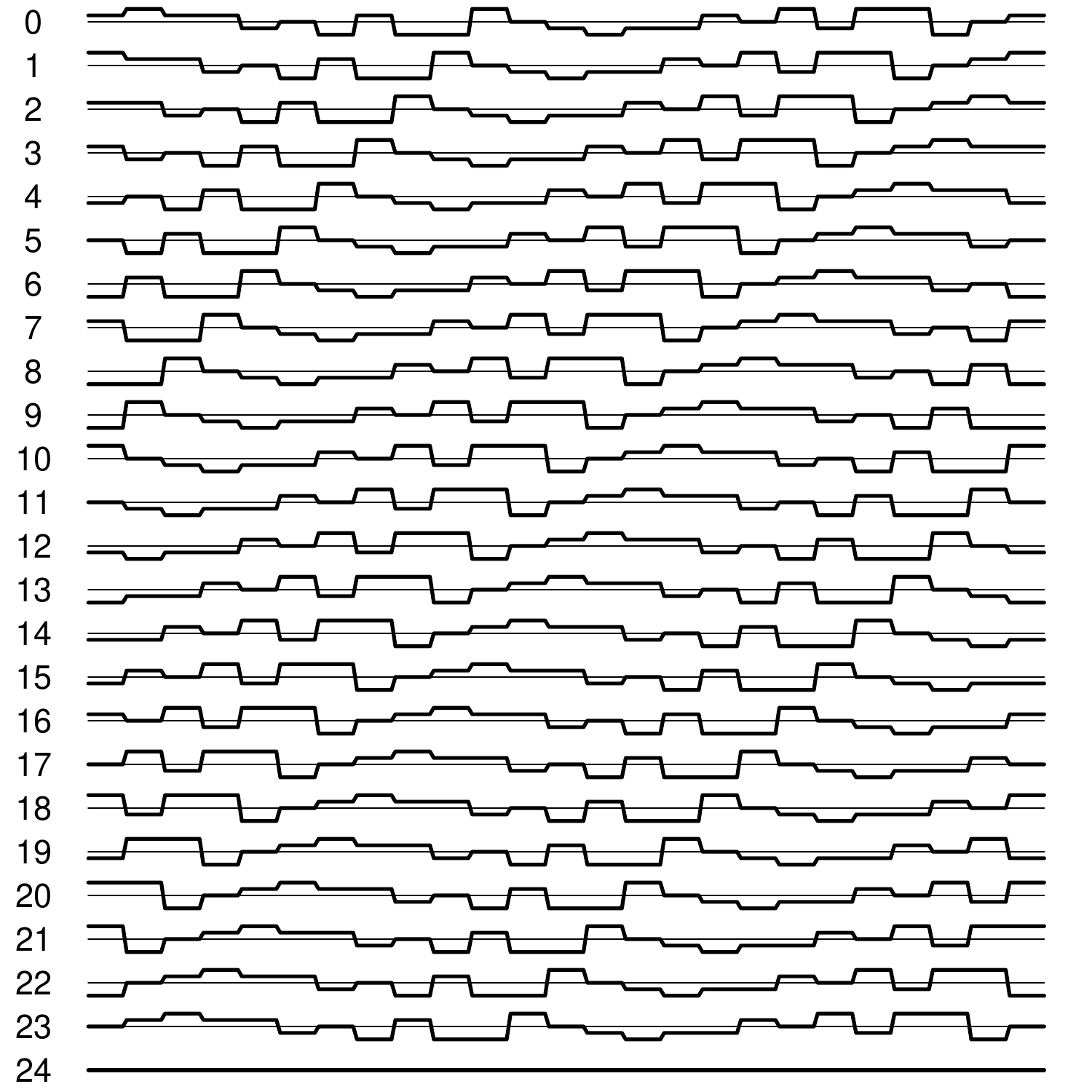}
\end{center}
\caption{The phases of a 25 baud alternating code. The cyclic
  nature of the code can be easily seen. One can also see property
  \ref{conj}, which states that each code is conjugate symmetric.}
\label{code25fig}
\end{figure}

\section{Discussion}

It is also possible to use truncated polyphase alternating codes in a
similar manner as binary phase alternating codes, in order to have smaller
number of scans.

Like binary alternating codes, also general $p$-nary codes made in
this way have the worst possible covariance structure, and need
randomization as described by \cite{lehtinen97}. The covariance
structures of $(p-1)$-nary codes have not been studied yet.

We have also done a complete search of codes with columns from $F_p$
for all numbers up to $p=15$. For numbers followed by a prime there is
indeed an unique code. On the other hand, for the numbers 8, 9, 14 and
15 there are no alternating code sets. These are the first composite
numbers not followed by a prime. This small search hints to the
possibility that there are no $p$-nary alternating codes of length $p$
formed from $F_p$, except when $p$ or $p+1$ is a prime.

\conclusions

Signal processing hardware today can easily generate arbitrary
waveforms, this includes the polyphase codes presented in this
paper. Because these codes also have a constant amplitude, there
is no transmission power trade-off compared to binary phase
codes. It should be pretty straightforward to modify existing
correlators to use these new codes. The main benefit of polyphase
alternating codes is the larger set of code lengths compared to binary
phase codes.


\begin{acknowledgements}
Support for J. Vierinen was provided by the Academy of Finland
(application number 213476, Finnish Programme for Centres of
Excellence in Research 2006-1011).
\end{acknowledgements}

\bibliographystyle{copernicus}
\bibliography{polyphaseAC}
\begin{table*}
\centering
\begin{tabular*}{\textwidth}{llllllllllll@{\extracolsep{\fill}}}
\hline
Length & $N_p$& $N_c$ & $b$ & Length & $N_p$ & $N_c$ & $b$& Length & $N_p$& $N_c$ & $b$ \\
\hline
$   2$ & $   2$ & $1$ & $1$ & $  88$ & $  88$ & $1$ & $ \downarrow $ & $ 226$ & $ 226$ & $1$ & $ \downarrow $    \\ 
$   3$ & $   3$ & $1$ & $2$ & $  89$ & $  89$ & $40$ & $3$ & $ 227$ & $ 227$ & $112$ & $2$   \\ 
$   4$ & $   2$ & $1$ & $1,1$ & $  96$ & $  96$ & $1$ & $ \downarrow $ & $ 228$ & $ 228$ & $1$ & $ \downarrow $    \\ 
$   4$ & $   4$ & $1$ & $ \downarrow $ & $  97$ & $  97$ & $32$ & $5$ & $ 229$ & $ 229$ & $72$ & $6$   \\ 
$   5$ & $   5$ & $2$ & $2$ & $ 100$ & $ 100$ & $1$ & $ \downarrow $ & $ 232$ & $ 232$ & $1$ & $ \downarrow $    \\ 
$   6$ & $   6$ & $1$ & $ \downarrow $ & $ 101$ & $ 101$ & $40$ & $2$ & $ 233$ & $ 233$ & $112$ & $3$   \\ 
$   7$ & $   7$ & $2$ & $3$ & $ 102$ & $ 102$ & $1$ & $ \downarrow $ & $ 238$ & $ 238$ & $1$ & $ \downarrow $    \\ 
$   8$ & $   2$ & $2$ & $1,0,1$ & $ 103$ & $ 103$ & $32$ & $5$ & $ 239$ & $ 239$ & $96$ & $7$   \\ 
$   9$ & $   3$ & $2$ & $1,1$ & $ 106$ & $ 106$ & $1$ & $ \downarrow $ & $ 240$ & $ 240$ & $1$ & $ \downarrow $    \\ 
$  10$ & $  10$ & $1$ & $ \downarrow $ & $ 107$ & $ 107$ & $52$ & $2$ & $ 241$ & $ 241$ & $64$ & $7$   \\ 
$  11$ & $  11$ & $4$ & $2$ & $ 108$ & $ 108$ & $1$ & $ \downarrow $ & $ 243$ & $   3$ & $22$ & $2,0,0,0,1$   \\ 
$  12$ & $  12$ & $1$ & $ \downarrow $ & $ 109$ & $ 109$ & $36$ & $6$ & $ 250$ & $ 250$ & $1$ & $ \downarrow $    \\ 
$  13$ & $  13$ & $4$ & $2$ & $ 112$ & $ 112$ & $1$ & $ \downarrow $ & $ 251$ & $ 251$ & $100$ & $6$   \\ 
$  16$ & $   2$ & $2$ & $1,0,0,1$ & $ 113$ & $ 113$ & $48$ & $3$ & $ 256$ & $   2$ & $16$ & $1,0,0,0,1,1,1,0$   \\ 
$  16$ & $  16$ & $1$ & $ \downarrow $ & $ 121$ & $  11$ & $16$ & $3,1$ & $ 256$ & $ 256$ & $1$ & $ \downarrow $    \\ 
$  17$ & $  17$ & $8$ & $3$ & $ 125$ & $   5$ & $20$ & $2,0,1$ & $ 257$ & $ 257$ & $128$ & $3$   \\ 
$  18$ & $  18$ & $1$ & $ \downarrow $ & $ 126$ & $ 126$ & $1$ & $ \downarrow $ & $ 262$ & $ 262$ & $1$ & $ \downarrow $    \\ 
$  19$ & $  19$ & $6$ & $2$ & $ 127$ & $ 127$ & $36$ & $3$ & $ 263$ & $ 263$ & $130$ & $5$   \\ 
$  22$ & $  22$ & $1$ & $ \downarrow $ & $ 128$ & $   2$ & $18$ & $1,0,0,0,0,0,1$ & $ 268$ & $ 268$ & $1$ & $ \downarrow $    \\ 
$  23$ & $  23$ & $10$ & $5$ & $ 130$ & $ 130$ & $1$ & $ \downarrow $ & $ 269$ & $ 269$ & $132$ & $2$   \\ 
$  25$ & $   5$ & $4$ & $2,2$ & $ 131$ & $ 131$ & $48$ & $2$ & $ 270$ & $ 270$ & $1$ & $ \downarrow $    \\ 
$  27$ & $   3$ & $4$ & $2,0,1$ & $ 136$ & $ 136$ & $1$ & $ \downarrow $ & $ 271$ & $ 271$ & $72$ & $6$   \\ 
$  28$ & $  28$ & $1$ & $ \downarrow $ & $ 137$ & $ 137$ & $64$ & $3$ & $ 276$ & $ 276$ & $1$ & $ \downarrow $    \\ 
$  29$ & $  29$ & $12$ & $2$ & $ 138$ & $ 138$ & $1$ & $ \downarrow $ & $ 277$ & $ 277$ & $88$ & $5$   \\ 
$  30$ & $  30$ & $1$ & $ \downarrow $ & $ 139$ & $ 139$ & $44$ & $2$ & $ 280$ & $ 280$ & $1$ & $ \downarrow $    \\ 
$  31$ & $  31$ & $8$ & $3$ & $ 148$ & $ 148$ & $1$ & $ \downarrow $ & $ 281$ & $ 281$ & $96$ & $3$   \\ 
$  32$ & $   2$ & $6$ & $1,0,0,1,0$ & $ 149$ & $ 149$ & $72$ & $2$ & $ 282$ & $ 282$ & $1$ & $ \downarrow $    \\ 
$  36$ & $  36$ & $1$ & $ \downarrow $ & $ 150$ & $ 150$ & $1$ & $ \downarrow $ & $ 283$ & $ 283$ & $92$ & $3$   \\ 
$  37$ & $  37$ & $12$ & $2$ & $ 151$ & $ 151$ & $40$ & $6$ & $ 289$ & $  17$ & $48$ & $3,4$   \\ 
$  40$ & $  40$ & $1$ & $ \downarrow $ & $ 156$ & $ 156$ & $1$ & $ \downarrow $ & $ 292$ & $ 292$ & $1$ & $ \downarrow $    \\ 
$  41$ & $  41$ & $16$ & $6$ & $ 157$ & $ 157$ & $48$ & $5$ & $ 293$ & $ 293$ & $144$ & $2$   \\ 
$  42$ & $  42$ & $1$ & $ \downarrow $ & $ 162$ & $ 162$ & $1$ & $ \downarrow $ & $ 306$ & $ 306$ & $1$ & $ \downarrow $    \\ 
$  43$ & $  43$ & $12$ & $3$ & $ 163$ & $ 163$ & $54$ & $2$ & $ 307$ & $ 307$ & $96$ & $5$   \\ 
$  46$ & $  46$ & $1$ & $ \downarrow $ & $ 166$ & $ 166$ & $1$ & $ \downarrow $ & $ 310$ & $ 310$ & $1$ & $ \downarrow $    \\ 
$  47$ & $  47$ & $22$ & $5$ & $ 167$ & $ 167$ & $82$ & $5$ & $ 311$ & $ 311$ & $120$ & $17$   \\ 
$  49$ & $   7$ & $8$ & $2,2$ & $ 169$ & $  13$ & $24$ & $2,4$ & $ 312$ & $ 312$ & $1$ & $ \downarrow $    \\ 
$  52$ & $  52$ & $1$ & $ \downarrow $ & $ 172$ & $ 172$ & $1$ & $ \downarrow $ & $ 313$ & $ 313$ & $96$ & $10$   \\ 
$  53$ & $  53$ & $24$ & $2$ & $ 173$ & $ 173$ & $84$ & $2$ & $ 316$ & $ 316$ & $1$ & $ \downarrow $    \\ 
$  58$ & $  58$ & $1$ & $ \downarrow $ & $ 178$ & $ 178$ & $1$ & $ \downarrow $ & $ 317$ & $ 317$ & $156$ & $2$   \\ 
$  59$ & $  59$ & $28$ & $2$ & $ 179$ & $ 179$ & $88$ & $2$ & $ 330$ & $ 330$ & $1$ & $ \downarrow $    \\ 
$  60$ & $  60$ & $1$ & $ \downarrow $ & $ 180$ & $ 180$ & $1$ & $ \downarrow $ & $ 331$ & $ 331$ & $80$ & $3$   \\ 
$  61$ & $  61$ & $16$ & $2$ & $ 181$ & $ 181$ & $48$ & $2$ & $ 336$ & $ 336$ & $1$ & $ \downarrow $    \\ 
$  64$ & $   2$ & $6$ & $1,0,0,0,0,1$ & $ 190$ & $ 190$ & $1$ & $ \downarrow $ & $ 337$ & $ 337$ & $96$ & $10$   \\ 
$  66$ & $  66$ & $1$ & $ \downarrow $ & $ 191$ & $ 191$ & $72$ & $19$ & $ 343$ & $   7$ & $36$ & $3,0,1$   \\ 
$  67$ & $  67$ & $20$ & $2$ & $ 192$ & $ 192$ & $1$ & $ \downarrow $ & $ 346$ & $ 346$ & $1$ & $ \downarrow $    \\ 
$  70$ & $  70$ & $1$ & $ \downarrow $ & $ 193$ & $ 193$ & $64$ & $5$ & $ 347$ & $ 347$ & $172$ & $2$   \\ 
$  71$ & $  71$ & $24$ & $7$ & $ 196$ & $ 196$ & $1$ & $ \downarrow $ & $ 348$ & $ 348$ & $1$ & $ \downarrow $    \\ 
$  72$ & $  72$ & $1$ & $ \downarrow $ & $ 197$ & $ 197$ & $84$ & $2$ & $ 349$ & $ 349$ & $112$ & $2$   \\ 
$  73$ & $  73$ & $24$ & $5$ & $ 198$ & $ 198$ & $1$ & $ \downarrow $ & $ 352$ & $ 352$ & $1$ & $ \downarrow $    \\ 
$  78$ & $  78$ & $1$ & $ \downarrow $ & $ 199$ & $ 199$ & $60$ & $3$ & $ 353$ & $ 353$ & $160$ & $3$   \\ 
$  79$ & $  79$ & $24$ & $3$ & $ 210$ & $ 210$ & $1$ & $ \downarrow $ & $ 358$ & $ 358$ & $1$ & $ \downarrow $    \\ 
$  81$ & $   3$ & $8$ & $1,0,0,1$ & $ 211$ & $ 211$ & $48$ & $2$ & $ 359$ & $ 359$ & $178$ & $7$   \\ 
$  82$ & $  82$ & $1$ & $ \downarrow $ & $ 222$ & $ 222$ & $1$ & $ \downarrow $ & $ 361$ & $  19$ & $48$ & $4,4$   \\ 
$  83$ & $  83$ & $40$ & $2$ & $ 223$ & $ 223$ & $72$ & $3$ & $ 366$ & $ 366$ & $1$ & $ \downarrow $    \\ 
\hline
\end{tabular*}
\caption{Alternating codes with lengths up to 366, only one code for
  each number of phases. The number of phases is denoted by $N_p$, the
  number of different code sets $N_c$ and the generator coefficients
  are denoted by $b$. The $(p-1)$-nary codes are generated from the
  next consecutive prime.}
\label{coderes}
\end{table*}


\end{document}